\documentclass[a4paper]{article}
\usepackage{a4wide}
\usepackage[hyphens]{url}
\usepackage{siunitx,booktabs,subcaption}
\usepackage{tabularx}
\usepackage{authblk}
\usepackage{amssymb}
\usepackage{lineno}
\usepackage{hyperref}
\usepackage{xcolor} 
\usepackage[normalem]{ulem} 
\usepackage{multicol}
\usepackage{graphicx}
\usepackage{listings}

\usepackage{tikz}

\usetikzlibrary{shapes}
\usetikzlibrary{arrows}
\usetikzlibrary{automata}
\usetikzlibrary{positioning}
\usetikzlibrary{hobby}

\tikzstyle{int}=[draw, fill=blue!20, minimum size=1em, text width=1cm]
\tikzstyle{intr}=[draw, fill=red!20, minimum size=1em, text width=1cm]
\tikzstyle{init}=[pin edge={to-, thin, black}]

\usepackage{subcaption}

\hypersetup{
  colorlinks=true,
  linkbordercolor=white,
  citebordercolor=white,
  urlbordercolor=white,
  linkcolor=blue,
  citecolor=blue,
  urlcolor=blue
}
\usepackage[numbers]{natbib}
\graphicspath{{./figures/}}

\title{FEniCS Mechanics: A Package for Continuum Mechanics
  Simulations}

\author[1,$\ast$]{Miguel A. Rodriguez}
\author[1,2]{Christoph M. Augustin}
\author[1]{Shawn C. Shadden}

\affil[1]{Department of Mechanical Engineering, University of California, Berkeley, CA, USA}
\affil[2]{Gottfried Schatz Research Center, Biophysics, Medical University of Graz, Graz, Austria}
\affil[$\ast$]{Correspondance: Miguel A. Rodriguez, University of California,
Berkeley, CA 94720-1740, USA. miguelr@berkeley.edu}
\date{}

\begin{document}

\maketitle

\begin{abstract}
FEniCS Mechanics is a Python package to facilitate computational
mechanics simulations. The Python library \textsf{dolfin}, from the
FEniCS Project, is used to formulate
and numerically solve the problem in variational form. The
general balance laws from continuum mechanics are used to enable rapid
prototyping of different material laws. In addition to its generality,
FEniCS Mechanics also checks the input provided by users to ensure
that problem definitions are physically consistent. In
turn, this code enables simulations of custom mechanics problems to be more
accessible to those with limited programming or mechanics knowledge.

{\textbf{Keywords:} computational mechanics, fluid mechanics, solid mechanics,
  finite element method}

\end{abstract}

\section{Motivation and significance}
\label{sec:motivation}

A mathematical description of the kinematics of deformation and its relation to forces acting on and within a body is the basis of continuum mechanics.
The mathematical formulation is, in general, a nonlinear set of partial differential equations (PDEs) with corresponding initial and boundary conditions (defining an initial-boundary value problem, IBVP) for which analytical solutions can only be obtained in specialized cases.
Thus, computational mechanics is employed to numerically approximate solutions for these problems.
With increasing computational power and sophistication of material models, computational mechanics has become a crucial area of engineering design and research.

While various methods are used to discretize PDEs for
numerical solutions, the FEniCS Mechanics software described herein
utilizes the finite element method (FEM) for spatial discretization via tools
provided by the FEniCS package
\cite{fenics_project}.
There are various packages for FEM modeling~\cite{fea-software},
including commercial software such as Abaqus\cite{abaqus},
ANSYS\cite{ansys}, ADINA\cite{adina} or Comsol\cite{comsol}; and
open-source software including Deal II~\cite{deal.II},
FreeFem$++$\cite{freefem} and FEniCS. Each has unique
strengths~\cite{fea-compare}, and FEniCS was chosen because it is
open-source, widely-used, well-supported and has broad capabilities
that can be leveraged by the FEniCS Mechanics software developed
here. The versatility and performance of FEniCS is covered in The
FEniCS Book~\cite{fenicsbook}.

FEniCS was developed to solve problems that can be formulated in variational form.
IBVPs arising from continuum mechanics fit well within this framework.
To solve problems in FEniCS, the variational form of the PDEs, function spaces, element types, solver settings, etc., are specified through the development of scripts.
However, it is possible to formulate a range of continuum mechanics problems similarly (see documentation).
In particular, different types of problems can be described by changing the constitutive relationship, and through the utilization or interpretation of various terms of a generalized variational form.
This fact is the basis for FEniCS Mechanics.
While a particular continuum mechanics problem can be solved in FEniCS through the development of appropriate scripts, these scripts would typically be specific to the type of problem (e.g., solid, fluid) and properties of the material (e.g., constitutive equations, compressibility).
FEniCS Mechanics however provides a framework so that a variety of continuum mechanics problems can be considered through minor changes to a configuration file, or potentially ``minimally-invasive'' changes to scripts defining the generalized mechanics problem.
This provides an efficient framework to consider a variety of mechanics problems, or testing of modeling choices (e.g. types of materials).
This more streamlined approach also increases accessibility to computational mechanics simulations for users with minimal programming knowledge, while still maintaining a powerful and extensible open-source framework that can access the full array of FEM and solver capabilities provided through the FEniCS project.

In terms of specific capabilities, FEniCS Mechanics supports both
steady-state and time-dependent problems in a single domain, and the
user can choose from a list of implemented material models, or provide
their own, so long as the material is either elastic (stress depends
on the deformation gradient) or viscous (stress depends on the
velocity gradient). Furthermore, elastic materials can be specified as
compressible or incompressible, whereas all viscous models are assumed
to be incompressible. Discretization in time is currently handled by
single-step finite-difference schemes, including $\theta$-method for
first order systems and Newmark scheme for second order systems (see
documentation). The user is advised that stabilization methods have
not yet been implemented. Hence, the mesh used for each problem should
be chosen carefully to avoid instabilities during
computations. Lastly, our design facilitates the addition of
user-defined material models and changes to existing algorithms. A
description of how the FEniCS Mechanics package can be used to solve
computational mechanics problems is given in Section
\ref{sec:description}. This is followed by a demonstrative example in
Section \ref{sec:examples}.

\section{Software description}
\label{sec:description}

The description of the problem to be solved is defined through a
Python dictionary, referred to as \textsf{config}. FEniCS Mechanics
then parses this dictionary to define the variational
forms through the Unified Form Language (UFL) from the FEniCS Project
\cite{fenics_project,fenicsbook}, which are then used for matrix
assembly in order to obtain numerical solutions to the specified
problem.

\subsection{Software Architecture}
\label{subsec:architecture}

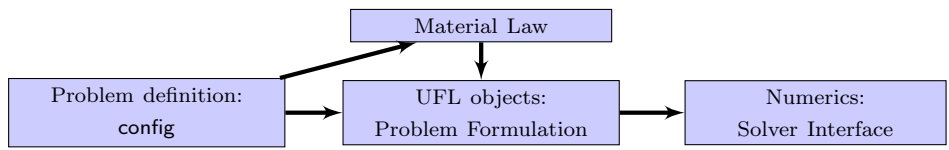
\begin{figure}[ht!]
  \centering
  \begin{tikzpicture}[auto, node distance=4.4cm,
      >=latex', align=center, scale=0.8]
    \node[int, text width=3.4cm] (config)
         {\footnotesize Problem definition:\\ \textsf{config}};
    \node[int, text width=3.4cm, right of=config] (problem1)
         {\footnotesize UFL objects: \\ Problem Formulation};
    \node[int, text width=3.2cm, above=0.5cm of problem1] (material)
         {\footnotesize Material Law};
    \node[int, text width=3.2cm, right of=problem1] (solver1)
         {\footnotesize Numerics: \\ Solver Interface};

    \path[ultra thick, ->] (config) edge (problem1);
    \path[ultra thick, ->] (problem1) edge (solver1);
    \path[ultra thick, ->] (config) edge (material);
    \path[ultra thick, ->] (material) edge (problem1);
  \end{tikzpicture}
  \caption{Diagram showing the flow of information within FEniCS
    Mechanics.}
  \label{fig:infoflow}
\end{figure}

Information flow in FEniCS Mechanics is shown in Figure~\ref{fig:infoflow}.
First, the user defines the problem by assigning values to various
keys in the \textsf{config} dictionary, e.g., material model, time
integration parameters, domain and mesh file. This \textsf{config}
dictionary is then provided to a problem class for instantiation.

All problem classes are intended to be derived from the
\textsf{BaseMechanicsProblem} class provided in FEniCS Mechanics,
including the three problem classes currently implemented and shown in
Figure~\ref{subfig:problems}. The base class provides methods that are
common to all mechanics problems, including the parsing of the
\textsf{config} dictionary.

\begin{figure}[ht!]
  \centering
  \begin{subfigure}[t]{0.47\textwidth}
    \centering
    \begin{tikzpicture}[auto, >=latex', align=center, scale=0.5]
      \node[intr] (base) {\footnotesize
        \textsf{BMP}}; 
      \node[intr, below left=0.5cm and 0.5cm of base] (mechanics)
           {\footnotesize \textsf{MP}}; 
      \node[intr, below right=0.5cm and 0.5cm of base]
      (solidmechanics) {\footnotesize \textsf{SMP}}; 
      \node[intr, below= 0.7cm of base] (fluidmechanics)
           {\footnotesize \textsf{FMP}};

      \path[ultra thick, ->] (base) edge (mechanics);
      \path[ultra thick, ->] (base) edge (solidmechanics);
      \path[ultra thick, ->] (base) edge (fluidmechanics);
    \end{tikzpicture}
    \caption{\footnotesize Classes that define the variational
      problem using the UFL from FEniCS.}
    \label{subfig:problems}
  \end{subfigure}
  \hfill
  \begin{subfigure}[t]{0.47\textwidth}
    \centering
    \begin{tikzpicture}[auto, >=latex', align=center, scale=0.5]
      \node[intr] (nonlinearsolver) {\footnotesize
        \textsf{NVS}}; 
      \node[intr, below=0.7cm of nonlinearsolver] (basesolver)
           {\footnotesize \textsf{BMS}};
      \node[intr, below left= 1.7cm and 0.2cm of nonlinearsolver]
      (solidsolver) {\footnotesize \textsf{SMS}}; 
      \node[intr, above left=0.1cm and 0.8cm of basesolver] (solver)
           {\footnotesize \textsf{MBS}}; 
      \node[intr, below right=1.7cm and 0.2cm of nonlinearsolver]
      (fluidsolver) {\footnotesize \textsf{FMS}};

      \path[ultra thick, ->] (nonlinearsolver) edge (basesolver);
      \path[ultra thick, ->] (basesolver) edge (solidsolver);
      \path[ultra thick, ->] (basesolver) edge (fluidsolver);
    \end{tikzpicture}
    \caption{\footnotesize Classes that assemble the variational forms defined from
      \ref{subfig:problems}, and call the numerical solvers.}
    \label{subfig:solvers}
  \end{subfigure}
  \begin{subfigure}[t]{\textwidth}
    \centering
    \begin{tikzpicture}[auto, >=latex', align=center, scale=0.5]
      \node[intr] (elastic) {\footnotesize
        \textsf{EM}}; 
      \node[intr, below left=15pt and 15pt of elastic] (isotropic)
          {\footnotesize \textsf{IM}}; 
      \node[intr, below left=12pt and -2pt of isotropic] (linear)
           {\footnotesize \textsf{LIM}}; 
      \node[intr, below right=12pt and -2pt of isotropic]
          (neohookean) {\footnotesize \textsf{NHM}}; 
      \node[intr, below=30pt of isotropic] (demiray)
           {\footnotesize \textsf{DM}}; 

      \node[intr, below right=15pt and 15pt of elastic] (anisotropic)
           {\footnotesize \textsf{AM}}; 
      \node[intr, below left=12pt and -15pt of anisotropic] (fung)
           {\footnotesize \textsf{FM}}; 
      \node[intr, below=0.6cm of fung] (guccione)
           {\footnotesize \textsf{GM}}; 
      \node[intr, below right=12pt and -15pt of anisotropic]
           (holzapfel) {\footnotesize \textsf{HOM}}; 

      \path[ultra thick, ->] (elastic) edge (isotropic);
      \path[ultra thick, ->] (isotropic) edge (linear);
      \path[ultra thick, ->] (isotropic) edge (neohookean);
      \path[ultra thick, ->] (elastic) edge (anisotropic);
      \path[ultra thick, ->] (anisotropic) edge (fung);
      \path[ultra thick, ->] (fung) edge (guccione);
      \path[ultra thick, ->] (isotropic) edge (demiray);
      \path[ultra thick, ->] (anisotropic) edge (holzapfel);

      \node[intr, below right=10pt and 100pt of elastic] (fluid) {\footnotesize
        \textsf{F}}; 
      \node[intr, below=0.8cm of fluid] (newtonian)
           {\footnotesize \textsf{NF}}; 

      \path[ultra thick, ->] (fluid) edge (newtonian);
    \end{tikzpicture}
    \caption{\footnotesize Classes that define the constitutive equations for
      different elastic (left) and fluid (right) materials
      initiated in \textsf{MP}, \textsf{SMP}, or \textsf{FMP}.}
    \label{subfig:materials}
  \end{subfigure}\\
  \begin{subfigure}{\textwidth}
    \vspace*{12pt}
    \centering
    \footnotesize
    \begin{tabular}{|c|c||c|c|} \hline
      Abbrev. & Full Name & Abbrev. & Full Name
          \\ \hline\hline
      \textsf{BMP} & \textsf{BaseMechanicsProblem} &
          \textsf{IM} & \textsf{IsotropicMaterial} \\ \hline
      \textsf{MP} & \textsf{MechanicsProblem} &
          \textsf{LIM} & \textsf{LinearIsoMaterial} \\ \hline
      \textsf{FMP} & \textsf{FluidMechanicsProblem} &
          \textsf{DM} & \textsf{DemirayMaterial
            \cite{demiray_1972}} \\ \hline
      \textsf{SMP} & \textsf{SolidMechanicsProblem} &
          \textsf{NHM} & \textsf{NeoHookeMaterial} \\ \hline
      \textsf{MBS} & \textsf{MechanicsBlockSolver} &
          \textsf{AM} & \textsf{AnisotropicMaterial} \\ \hline
      \textsf{NVS} & \textsf{NonlinearVariationalSolver}$^*$ &
          \textsf{FM} & \textsf{FungMaterial}
          \cite{humphrey_1995} \\ \hline
      \textsf{BMS} &
      \textsf{BaseMechanicsSolver} &
          \textsf{GM} & \textsf{GuccioneMaterial}
          \cite{guccione_costa_mcculloch_1995} \\ \hline
      \textsf{SMS} & \textsf{SolidMechanicsSolver} &
          \textsf{HOM} &
          \textsf{HolzapfelOgdenMaterial}
            \cite{holzapfel_ogden_2009} \\ \hline
      \textsf{FMS} & \textsf{FluidMechanicsSolver} &
          \textsf{F} & \textsf{Fluid} \\ \hline
      \textsf{EM} & \textsf{ElasticMaterial} &
          \textsf{NF} & \textsf{NewtonianFluid} \\ \hline
    \end{tabular}
    \caption{\footnotesize Abbreviations used in the tree diagrams above.}
    \label{tab:abbreviations}
  \end{subfigure}
  \caption{Inheritance diagrams for all classes defined within
    FEniCS Mechanics, and their
    abbreviations. $^*$\textsf{NonlinearVariationalSolver} is defined
    in \textsf{dolfin} from the FEniCS project.
  }
  \label{fig:inheritance}
\end{figure}

Once \textsf{config} is parsed, the respective problem class --
currently \textsf{MechanicsProblem}, \textsf{SolidMechanicsProblem},
or \textsf{FluidMechanicsProblem} -- defines the variational equations
for the respective problem using the UFL. The
\textsf{MechanicsProblem} class defines the variational problem with
separate function spaces for vector and scalar valued field variables,
whereas the \textsf{SolidMechanicsProblem} and
\textsf{FluidMechanicsProblem} classes use the mixed function space
functionality of \textsf{dolfin}. The information pertaining to material
models in \textsf{config} is passed to separate classes defining
constitutive equations. This can be seen in Figure
\ref{fig:infoflow}.

With the variational equations defined through the UFL and stored as
member data of a problem object, a solver object is next created. The
three current solver classes are \textsf{MechanicsBlockSolver},
\textsf{SolidMechanicsSolver}, and \textsf{FluidMechanicsSolver}, and
are to be used with \textsf{MechanicsProblem},
\textsf{SolidMechanicsProblem}, and \textsf{FluidMechanicsProblem},
respectively. These solver objects have methods that use the UFL forms
from the problem objects to assemble the resulting linear algebraic
system at each iteration of a nonlinear solve. This is repeated for
each time step if the problem is time-dependent. Note that
\textsf{SolidMechanicsSolver} and \textsf{FluidMechanicsSolver} are a
subclass of the \textsf{NonlinearVariationalSolver} class from
\textsf{dolfin} through the \textsf{BaseMechanicsSolver} class,
while \textsf{MechanicsBlockSolver} is a stand-alone block solver
class based on the FEniCS Application, CBC-Block
(\url{https://bitbucket.org/fenics-apps/cbc.block}), as shown in
Figure~\ref{subfig:solvers}.

The user is expected to interact the most with the problem and solver
classes mentioned above. However, in addition to these, various
constitutive models have been implemented in a \textsf{materials}
sub-module within FEniCS Mechanics. These constitutive models and
their inheritance trees are shown in Figures \ref{subfig:materials} and
\ref{tab:abbreviations}.

\subsection{Software Functionalities}
\label{subsec:functionalities}

In order to facilitate problem definition and accessibility, FEniCS
Mechanics implements the following functionalities:
\begin{enumerate}
\item Key-value pairs provided in \textsf{config} and their
  combinations are checked for validity. This increases accessibility
  by making sure that invalid, or inconsistent, values in
  \textsf{config} are not used.
\item FEniCS Mechanics uses the problem specification provided in
  \textsf{config} to define the variational form using the UFL.
\item The variational form defined through the UFL is used to assemble
  the resulting linear systems and obtain a numerical solution to the
  problem.
\end{enumerate}
All three functionalities are demonstrated in the example of Section
\ref{sec:examples}.

\section{Illustrative Example}
\label{sec:examples}

\begin{figure}[ht!]
  \centering
  \begin{subfigure}[t]{0.6\textwidth}
    \centering
    \includegraphics[width=\textwidth]{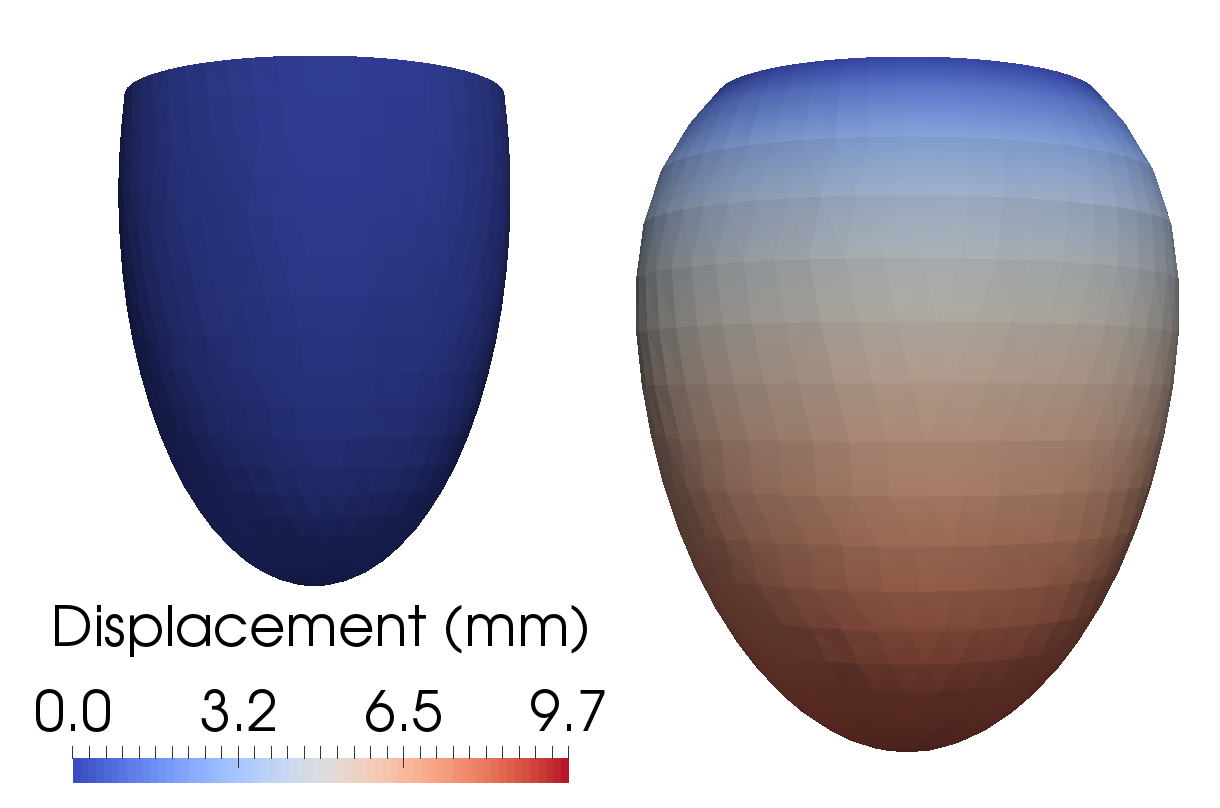}
    \caption{\footnotesize The unloaded (left) and loaded (right) configurations.}
    \label{subfig:disp}
  \end{subfigure}%
  \hfill
  \begin{subfigure}[t]{0.3\textwidth}
    \centering
    \includegraphics[width=\textwidth]{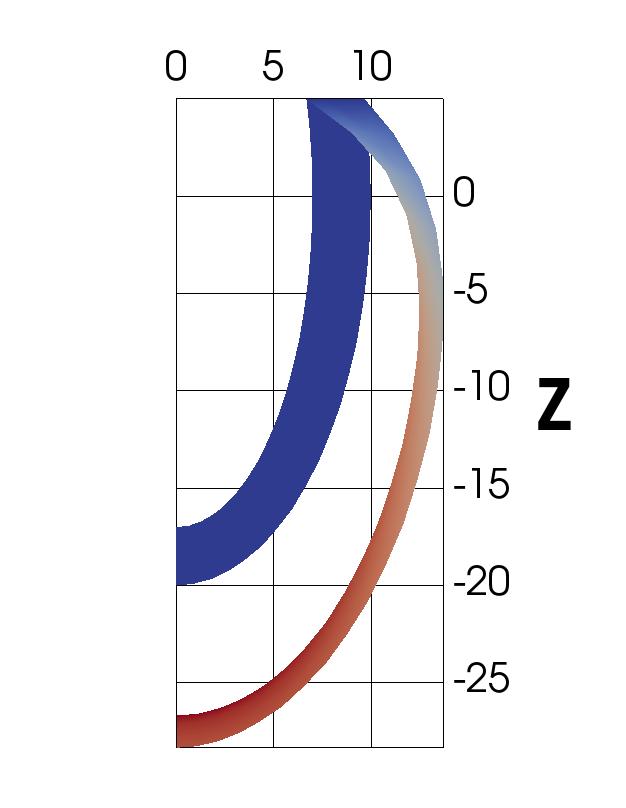}
    \caption{\footnotesize A sliced view of both configurations.}
    \label{subfig:slice}
  \end{subfigure}
  \caption{Displacement at the final loading step, $p = 10$ kPa, of
    the idealized left ventricle.}
  \label{fig:ellipsoid}
\end{figure}

Consider a truncated ellipsoid (Fig.~\ref{subfig:disp}, left) as
considered in Land et al.~\cite{land_et_al_2015}, which is used to
model an idealized left ventricle of the heart. This example also
serves as verification of FEniCS Mechanics by enabling direct
comparison with benchmark solutions presented in ~\cite{land_et_al_2015}.
The model geometry is mechanically loaded by
applying a 10 kPa pressure at the inner wall, and the base (top) plane
is fixed in all directions. For stability, the pressure is ramped from
0 to 10 kPa in 100 steps. Rather than manually writing a for-loop to
solve a quasi-static problem for each of the 100 loading steps, we
take advantage of the fact that FEniCS Mechanics supports
time-dependent problems. Therefore, the applied pressure is defined as
$\bar{p}(t) = 10t$ kPa for $t \in [0,1]$, and a time step of $\Delta t
= 0.01$ is used. To ensure the problem solved at each time step is
quasi-static, the density of the material is set to zero to exclude
the inertial term from the weak form for the balance of
momentum. Passive material behavior is characterized by an
incompressible transversely-isotropic constitutive equation as
described in Guccione~et~al.~\cite{guccione_costa_mcculloch_1995}.
This constitutive equation is a subclass of the
  \textsf{FungMaterial} model (Fig.~\ref{subfig:materials}) that is
  provided and is discussed in detail by Humphrey
  \cite{humphrey_1995}. Constitutive equations of this form were
  developed to model soft biological tissues, such as arterial and
  cardiac tissue, and hence was chosen for the comparative study of
  various cardiac mechanics software by Land et
  al.~\cite{land_et_al_2015}. The interested reader is referred to
  \cite{humphrey_1995, guccione_costa_mcculloch_1995,
    holzapfel_ogden_2009, costa_holmes_mcculloch_2001} for further
  details and the references therein. The \textsf{config} dictionary
for this problem is shown in Listing~\ref{code1}.
\begin{lstlisting}[language=python,frame=single,caption=Python code,label=code1]
import fenicsmechanics as fm

# Material model and parameters
mat_dict = {'const_eqn': 'guccione', 'type': 'elastic',
    'incompressible': True, 'density': 0.0,
    'bt': 1.0, 'bf': 1.0, 'bfs': 1.0, 'C': 10.0,
    'fibers': {
        'fiber_files': ['n1-p0.xml.gz',
                        'n2-p0.xml.gz'],
        'fiber_names': ['n1', 'n2'], 'elementwise': True}}

# Mesh file names
mesh_dict = {'mesh_file': 'mesh.xml.gz',
             'boundaries': 'boundaries.xml.gz'}

# Time integration parameters, BCs, and polynomial degree.
formulation_dict = {
    'time':{'dt': 0.01, 'interval': [0., 1.]},
    'element': 'p2-p1',
    'domain': 'lagrangian',
    'bcs':{
        'dirichlet': {
            'displacement': [[0., 0., 0.]],
            'regions': [10], # Integer ID for base plane
            },
        'neumann': {
            'regions': [20], # Integer ID for inner surface
            'types': ['pressure'],
            'values': ['10.0*t']}}}

# Combine above dictionaries into one.
config = {'material': mat_dict, 'mesh': mesh_dict,
          'formulation': formulation_dict}

# Create problem and solver objects.
problem = fm.SolidMechanicsProblem(config)
solver = fm.SolidMechanicsSolver(problem,
                                 fname_disp='displacement.pvd')

# Numerically solve the problem.
solver.full_solve()
\end{lstlisting}

The original (unloaded) and final (loaded) configurations of the
idealized left ventricle using a mesh element size of 1000
  $\mu$m are shown in Figure \ref{fig:ellipsoid}. The endocardial
(inner wall) apex is at $(0,0,-17)$ mm in the unloaded
configuration. It can be seen that in the loaded configuration, a
displacement of about 9.7 mm in the negative $z$
direction gives $(0,0,-26.7)$ as its final
position. This is in broad agreement with the results of the
participating groups of the benchmark paper, see Figure 6 of
Land~et~al.~\cite{land_et_al_2015}. The number of degrees-of-freedom (DOF)
for displacement and pressure, and the resulting endocardial and epicardial
apex locations are given for different mesh element sizes in Table
\ref{tab:convergence}.
These results are provided for direct comparison with Figure 6 in Land et
  al.~\cite{land_et_al_2015}.

\begin{table}[ht!]
  \centering
  \begin{tabular}{|c|c|c|c|c|} \cline{4-5}
    \multicolumn{3}{l}{} & \multicolumn{2}{|c|}{Final Apex Location ($z$, mm)} \\ \hline
    Mesh Size ($\mu$m) & DOF($\mathbf{u}$) & DOF($p$) & Endocardium &
    Epicardium \\ \hline\hline
    2000 &    6264 &   362 & -26.85 & -28.09 \\ \hline
    1000 &   63630 &  3084 & -26.68 & -28.34 \\ \hline
     500 &  475176 & 21504 & -26.67 & -28.33 \\ \hline
     300 & 2156805 & 94622 & -26.67 & -28.33 \\ \hline
  \end{tabular}
  \caption{Convergence study results for the ellipsoid
      problem. \textsf{P2-P1} elements were used for all
      simulations. Endocardial and epicardial apex locations are in agreement with Figure 6
      in Land et al.~\cite{land_et_al_2015}.}
  \label{tab:convergence}
\end{table}

\section{Impact and Conclusions}
\label{sec:impact}

Benefits of FEniCS Mechanics stem from the fact that it is built on
top of the FEniCS Project. Hence no additional installations are
required other than the optional CBC-Block application, as mentioned
in Section \ref{subsec:architecture}. FEniCS provides an interface to
state-of-the-art linear solvers and preconditioners from freely
available third-party libraries, e.g. PETSc, HYPRE, and Eigen. This is
an advantage not found in many other open source projects, which
depend on their own solver packages. Another main benefit of using
FEniCS as the backbone is the inheritance of parallelization. Problems
formulated with FEniCS Mechanics can be run in parallel, given that
FEniCS is installed with MPI support. Due to its design as a Python
dictionary the simulations are executed in a similar fashion on a
high-performance computing cluster as on a regular desktop computer.

A key benefit of FEniCS Mechanics, which distinguishes it from
standalone FEniCS, is the reduction of programming skills and
theoretical numerical knowledge required from the user. Several common
material models have been implemented and the current problem
classes span a wide range of possible applications. Altogether, this
facilitates access to experimentation and comparison of solutions from
differing problem definitions. The names of dictionary keys and values
have been chosen to be intuitive to minimize the gap between a user
and the simulation they wish to run. This also enables FEniCS
Mechanics to raise exceptions when erroneous or inconsistent values are
provided by a user, as described in Section \ref{subsec:functionalities}.

While a simplistic interface was maintained in FEniCS Mechanics to
facilitate running simulations, advanced users can take full advantage
of additional tools provided by the FEniCS Project for altering
problem definitions, such as, but not limited to,
providing their own constitutive equations for rapid
prototyping. Advanced users can also more finely tune the solver and
preconditioner parameters used, and perform post-processing tasks
within the same script.

\section*{Acknowledgements} \label{sec:acknowledgements}
This work was supported in part from the NSF award 1663747, the NSF GRFP, and a Marie Sklodowska-Curie fellowship (GA No 750835) to CA by the European Union's Horizon 2020 research and innovation program .



\bibliographystyle{plainnat}
\bibliography{articles}

\begin{thebibliography}{16}
\providecommand{\natexlab}[1]{#1}
\providecommand{\url}[1]{\texttt{#1}}
\expandafter\ifx\csname urlstyle\endcsname\relax
  \providecommand{\doi}[1]{doi: #1}\else
  \providecommand{\doi}{doi: \begingroup \urlstyle{rm}\Url}\fi

\bibitem[dea()]{deal.II}
The deal.ii finite element library.
\newblock URL \url{https://dealii.org}.

\bibitem[fen()]{fenics_project}
Fenics project.
\newblock \url{https://fenicsproject.org/}.
\newblock Accessed: Nov. 2016.

\bibitem[fre()]{freefem}
Freefem++.
\newblock URL \url{http://www.freefem.org}.

\bibitem[Abaqus(2018)]{abaqus}
Abaqus.
\newblock \emph{version 2018}.
\newblock Simulia, Providence, Rhode Island, 2018.
\newblock URL
  \url{https://www.3ds.com/products-services/simulia/products/abaqus/}.

\bibitem[ADINA(2018)]{adina}
ADINA.
\newblock \emph{version 9.4.3}.
\newblock ADINA R\&D Inc., Watertown, Massachusetts, 2018.
\newblock URL \url{http://www.adina.com/}.

\bibitem[ANSYS(2018)]{ansys}
ANSYS.
\newblock \emph{version 19.2}.
\newblock ANSYS, Inc., Canonsburg, Pennsylvania, 2018.
\newblock URL \url{https://www.ansys.com/}.

\bibitem[{COMSOL Multiphysics}(2018)]{comsol}
{COMSOL Multiphysics}.
\newblock \emph{version 5.3}.
\newblock COMSOL AB, Stockholm, Sweden, 2018.
\newblock URL \url{https://www.comsol.com/}.

\bibitem[Costa et~al.(2001)Costa, Holmes, and
  Mcculloch]{costa_holmes_mcculloch_2001}
Kevin~D. Costa, Jeffrey~W. Holmes, and Andrew~D. Mcculloch.
\newblock Modelling cardiac mechanical properties in three dimensions.
\newblock \emph{Philosophical Transactions of the Royal Society A:
  Mathematical, Physical and Engineering Sciences}, 359\penalty0
  (1783):\penalty0 1233--1250, 2001.
\newblock \doi{10.1098/rsta.2001.0828}.

\bibitem[Demiray(1972)]{demiray_1972}
Hilmi Demiray.
\newblock A note on the elasticity of soft biological tissues.
\newblock \emph{Journal of Biomechanics}, 5\penalty0 (3):\penalty0 309--311,
  1972.
\newblock \doi{10.1016/0021-9290(72)90047-4}.

\bibitem[Guccione et~al.(1995)Guccione, Costa, and
  Mcculloch]{guccione_costa_mcculloch_1995}
Julius~M. Guccione, Kevin~D. Costa, and Andrew~D. Mcculloch.
\newblock Finite element stress analysis of left ventricular mechanics in the
  beating dog heart.
\newblock \emph{Journal of Biomechanics}, 28\penalty0 (10):\penalty0
  1167--1177, 1995.
\newblock \doi{10.1016/0021-9290(94)00174-3}.

\bibitem[Holzapfel and Ogden(2009)]{holzapfel_ogden_2009}
Gerhard~A. Holzapfel and Ray~W. Ogden.
\newblock Constitutive modelling of passive myocardium: a structurally based
  framework for material characterization.
\newblock \emph{Philosophical Transactions of the Royal Society A:
  Mathematical, Physical and Engineering Sciences}, 367\penalty0
  (1902):\penalty0 3445--3475, Mar 2009.
\newblock \doi{10.1098/rsta.2009.0091}.

\bibitem[Humphrey(1995)]{humphrey_1995}
Jay~D. Humphrey.
\newblock Mechanics of the arterial wall: Review and directions.
\newblock \emph{Critical Reviews in Biomedical Engineering}, 23\penalty0
  (1-2):\penalty0 1--162, 1995.
\newblock \doi{10.1615/critrevbiomedeng.v23.i1-2.10}.

\bibitem[Ladutenko()]{fea-compare}
K.~Ladutenko.
\newblock Fea-compare.
\newblock URL \url{https://github.com/kostyfisik/FEA-compare}.

\bibitem[Land et~al.(2015)Land, Gurev, Arens, Augustin, Baron, Blake, Bradley,
  Castro, Crozier, Favino, and et~al.]{land_et_al_2015}
Sander Land, Viatcheslav Gurev, Sander Arens, Christoph~M. Augustin, Lukas
  Baron, Robert Blake, Chris Bradley, Sebastian Castro, Andrew Crozier, Marco
  Favino, and et~al.
\newblock Verification of cardiac mechanics software: benchmark problems and
  solutions for testing active and passive material behaviour.
\newblock \emph{Proceedings of the Royal Society A: Mathematical, Physical and
  Engineering Science}, 471\penalty0 (2184):\penalty0 20150641, Aug 2015.
\newblock \doi{10.1098/rspa.2015.0641}.

\bibitem[Logg et~al.(2016)Logg, Mardal, and Wells]{fenicsbook}
Anders Logg, Kent-Andre Mardal, and Garth Wells.
\newblock \emph{Automated Solution of Differential Equations by the Finite
  Element Method: The FEniCS Book}.
\newblock Springer Berlin, 2016.

\bibitem[Wikipedia()]{fea-software}
Wikipedia.
\newblock List of finite element software packages.
\newblock URL
  \url{https://en.wikipedia.org/wiki/List_of_finite_element_software_packages}.

\end{thebibliography}

\section*{Required Metadata}

\section*{Current code version}


\begin{table}[!h]
\begin{tabular}{|l|p{6.5cm}|p{6.5cm}|}
\hline
\textbf{Nr.} & \textbf{Code metadata description} & \textbf{Please fill in this column} \\
\hline
C1 & Current code version & v1.0 \\
\hline
C2 & Permanent link to code/repository used for this code version & \url{https://gitlab.com/ShaddenLab/fenicsmechanics} \\
\hline
C3 & Legal Code License   & BSD-3-clause \\
\hline
C4 & Code versioning system used & git \\
\hline
C5 & Software code languages, tools, and services used & Python \\
\hline
C6 & Compilation requirements, operating environments \& dependencies
& FEniCS Project, Version 2016.1.0 and up \\
\hline
C7 & If available Link to developer documentation/manual & \url{https://shaddenlab.gitlab.io/fenicsmechanics} \\
\hline
C8 & Support email for questions & \href{mailto:miguelr@berkeley.edu}{miguelr@berkeley.edu}\\
\hline
\end{tabular}
\caption{Code metadata}
\label{}
\end{table}




\end{document}